# Frustrated vacancy ordering creates novel quantum properties in Kutinaite, $Ag_6Cu_{14.4}As_7$

Authors: Kim-Khuong Huynh*[a], Rasmus Baden Stubkjær[a], Ventrapati Pavankumar[b], Emilie Skytte Vosegaard[a], Karl Omer Rimon Juul[a], Kasper Rasmussen Borup[a], Yong P. Chen[c], and Bo Brummerstedt Iversen*[a]

Affiliations:

[a]Center for Integrated Materials Research, Department of Chemistry and iNANO, Aarhus University, Langelandsgade 140, 8000 Aarhus (Denmark)

[b]Department of Chemistry, SRM University Amravati-AP, Mangalagiri 522240 AndhraPradesh, India

[c]Center for Integrated Materials Research, Department of Physics and iNANO, Aarhus University, Langelandsgade 140, 8000 Aarhus (Denmark)

*Corresponding authors: hkkhuong@chem.au.dk, bo@chem.au.dk

Abstract:

Ideal crystals are fully ordered, but real-world crystals always contain defects breaking translational symmetry. Random defects in crystals have important implications and they e.g. provide the foundation for semiconductor-based electronic devices. Structurally correlated defects introduce an additional level of complexity, which may lead to novel materials properties, but rationalization of relations between correlated disorder and the emergent material properties are very rare. Here we report that the defect structure of the mineral Kutinaite, $Ag_6Cu_{14.4}As_7$, exhibits unprecedented metallic diamagnetism, a hallmark of non-trivial electronic states that require delicate symmetrical protection. Using a combination of X-ray scattering methodologies, simulations, and physical property measurements, we deduced and verified subtle frustrated vacancy ordering of the Cu sublattice when cooling crystals below ~300 K. The vacancy frustration in Kutinaite leads to unique quantum properties, and our study calls for a reconsideration of the role of vacancies as quasi-chemical species in crystals.

# Introduction

Ideal crystals have perfect translational symmetry, allowing a complete description of the atomic structure using a unit cell. This simplifies descriptions of crystalline solids from ~$10^{23}$ individual atoms to maximum thousands of atoms. For a century the ideal periodic description of crystals has been a cornerstone of natural science. However, real world crystals always contain imperfections. Point defects, such as lattice vacancies, substitutional impurities or interstitial impurities, are in thermodynamic equilibrium with the host crystal [1,2], and their inevitable existence makes real crystals more complex than their idealized mathematical representations. Random point defects are essential for the properties of materials, and as an example the engineered bipolar electric conductivity of semiconductors by impurity doping underlies every electronic device. More fundamental than just bringing extrinsic properties to a preexisting parent phase, it is increasingly realized that defects can also intrinsically participate as an integral composition in the creation of entirely new materials with emergent properties [3]. This occurs when the defects establish correlated disorder due to some underlying local chemical interactions [4-6].

Many materials have atomic vacancy defects to achieve specific chemical compositions. A high concentration of vacancies can result in strongly correlated vacancy order (VO) occurring on top of the otherwise stable average structure. The significance of VO in crystalline materials was e.g. elaborated in thermoelectric half-Heusler $Nb_{0.83}CoSb$ crystals leading to a reduction of thermal conductivity by ~15% [7]. In Prussian blue analogue battery materials, the local VO significantly affects the electrochemical behaviors [8]. In $Eu_2ZnSb_2$, different VOs lead to distinct electronic topological phases and improved thermoelectric performances [9,10]. The richness of VOs in iron chalcogenides underlies their myriads of magnetic and superconducting phases [11,12]. In these examples, vacancies play the role of a special "chemical species" that impart a "superorder" on top of the original crystal structure [5,6].

The title material Kutinaite, $Ag_6Cu_{14.4}As_7$ [Fig. 1], which is the only known arsenide mineral of both silver and copper [13], circumvents the constraint of charge balance by including vacancies in its structure. Kutinaite adopts the complex face-centered cubic structure of $M_6Ni_{16}Si_7$ ($M$ is Mg or a transition metal) [13,14], where 6 Ag(+I) and 7 As(-III) fully occupy the $M$ and Si sites, respectively. However, a full occupancy of 16 Cu sites with Cu(+I) would lead to a surplus of positive charges. The mineral restores charge balance by emptying ~20% of the Cu2 sites, resulting in the chemical formula $Ag_6Cu_{14.4}As_7$. The exact nature of vacancy formation has profound consequences for the physical properties of the material.

This study concerns the vacancies inherent in the structure of Kutinaite and its unprecedented metallic properties. The vacancies distribute randomly at high temperatures, but at $T_{vo} \approx 300$ K they attain a superstructure ordering. We combine single crystal (SCXRD) and powder X-ray diffraction (PXRD), three-dimensional difference pair-distribution function (3D-ΔPDF) [15,16], and Monte-Carlo (MC) simulation [7], to uncover ordering rules that cannot define a unique vacancy configuration, and thus the superstructure is frustrated. High residual entropy and metastable behaviors in transport properties appear because of this frustration.

Other materials with this structure type exhibit conventional metallic properties [14], but the defect Kutinaite exhibits rare characteristics found only in clean crystals of non-trivial metals [17-19]. Kutinaite is a metal with strong electronic diamagnetism originating from the dynamics of extraordinarily fast charge carriers according to Landau-Peierls (LP) theory [20]. A Dirac-like non-trivial electronic structure is usually the source of these highly mobile carriers, as verified among the few known LP diamagnets. The magnitude of the magnetic susceptibility in Kutinaite is comparable to that of the record holder Bismuth. The remarkable coexistence of vacancies and strong diamagnetism provides new perspectives on the role of correlated disorders.

# Results
## Structure and vacancy ordering
### The average structure of Kutinaite

The SCXRD average structure of Kutinaite [Fig. 1] adopts the cubic $Fm\bar{3}m$ space group at 25 K-350 K with unit cell parameter $a$ = 11.78 Å (Supplementary Material (SM) I.i) and atomic positions (Wyckoff site in parenthesis): Ag1 (24e), As1 (4b), As2 (24d), Cu1 (32f) and Cu2 (32f). The elongated anisotropic atomic displacement parameters (ADPs) of As2 [Fig. 1a insert] are explained by anharmonic vibrations (SM I.ii), aligning with the observation of a large Boson peak in the specific heat capacity of Kutinaite at 5 K-100 K (SM IV). The refined Cu2 occupancy of ∼ 0.8 corresponds to 20% vacancies at all temperatures (SM I.iii). Cu2 occupies the vertices of $Cu_4$ tetrahedra [Fig. 1b], and an occupancy of 0.8 suggests that 4/5 of the tetrahedra are missing a vertex ($Cu_3$), whereas 1/5 of them are fully intact ($Cu_4$). The refined stoichiometry of Kutinaite is then $Ag_6Cu_{14.4}As_7$. An full occupancy, $Ag_6Cu_{16}As_7$, would lead to overall charge imbalance of +1, assuming the valences Ag(+I), Cu(+I) and As(-III)[14], while our observed stoichiometry has an imbalance of −0.6. The charge balance can be restored with a small amount of $Cu^{2+}$ ions [13]. The SCXRD average structure solution reproduces PXRD patterns at 240 K-400 K [Fig. 2(a)], and all reflections are indexed by the $Fm\bar{3}m$ space group with Rietveld refinement yielding $R_{wp} \leq 6\%$ (Fig. S4). Small impurities (<1 wt%) of Ag and $As_4O_6$ were identified in the powder material.

## Vacancy ordering
### X-ray scattering

The average structure of Kutinaite is robust at all temperatures. However, $T$-dependent SCXRD and PXRD reveal a reversible dis-/appearance of weak reflections when warming/cooling the material across a broad transitional window from 250 K to 350 K [Fig. 2]. These peaks are forbidden by $Fm\bar{3}m$ systematic absence rules, but modelling the low-$T$ average structure using a previously suggested less symmetrical space group ($Pm\bar{3}m$) [13] led to non-positive definite ADPs, despite including all peaks. The F-forbidden reflections are very weak, and at 25 K their mean SCXRD intensity and mean I/σ are 0.8 and 0.4 as compared to 25.7 and 13.2 for the F-allowed reflections. However, their sharpness implies a rather long-range correlation. We propose that the partially occupied Cu2 sublattice goes through a superstructure vacancy ordering (VO) transition at $T < T_{vo} \approx$ 300 K, and coherent scattering from the VO sublattice is the cause of the F-forbidden reflections, while retaining the $Fm\bar{3}m$ average structure. At $T \gg T_{\text{vo}}$, the thermal energy outweighs the correlation so that Cu2 vacancies distribute randomly and only the $Fm\bar{3}m$ reflections exist.

*Monte-Carlo simulation*
The nature of the VO was explored using MC simulations with Rietveld refinements as a reference [7]. We used a supercell of $5 \times 5 \times 5$ unit cells ($Fm\bar{3}m$) with 20% empty Cu2 sites as refined from SCXRD. The most energetically favourable configuration of Cu2 vacancies has them spaced as far apart as possible. We assign $J_i$ as the (positive) interaction energy between a Cu2 vacancy $v$ and each of the $N_i$ vacancies in its $i^{\text{th}}$ nearest neighbour (NN) sphere and introduce the dependence on distance as $J_1 = 2J_2 = 4J_3 = 10J_4 = 10^3 J_5$. The total energy $E_{\text{MC}}$ of a configuration considering up to $k^{\text{th}}$ NNs is then formulated as a summation over all $v$ vacancies;

$$E_{MC} = \sum_{v} \sum_{i}^{k} J_i N_i .$$

$E_{\text{MC}}$ converges to 0 for the simulations including penalties from 1st to 4th NNs meaning that a vacancy configuration can be achieved where the NN vacancy is further away than the 4th NN. However, if we include the 5th NN term the energy becomes positive and the restrictions cannot be met for all vacancies, meaning that some vacancy pairs will be placed at a distance corresponding to the 5th NN interatomic distance of 6.2 Å. Rietveld refinement of simulated configurations to the experimental PXRD data show that the $k = 4$ model reproduce all low-$T$ reflections [Fig. 3a].

*3D Difference Pair Distribution Function*
Correlated disorder results in diffuse scattering superimposed on the Bragg peak intensity from the average structure and can be analyzed using the 3D-ΔPDF technique [15,16]. The F-forbidden reflections arise from a strong inter-vacancy correlation and 3D-ΔPDF-like maps were constructed by Fourier transformation of the F-forbidden peaks fulfilling at least one of the criteria $h + k \neq 2n$, $h + l \neq 2n'$, or $k + l \neq 2n''$, hereby assuming that the effects of VO are contained in the additional peaks.

If we put a vacancy at the origin of the Fourier map [Fig. 3b], the appearance of strong negative 1$^{st}$ NN correlations in the $z = 0$ plane indicates that only one vacancy exists in each Cu2 tetrahedron. A similar signature is seen for the 2$^{nd}$ NN. The features to the 3$^{rd}$ and 4$^{th}$ NNs are weakly negative, and they can be understood by also considering the positive 5$^{th}$ NN correlation at interatomic vectors of 6.2 Å in both $z = 0$ and $z = 0.15$ planes. This reveals the preferred vacancy distance and explains the weakly negative features of the 3$^{rd}$ and 4$^{th}$ NNs as they are closer to another vacancy. Given a specific vacancy position in a tetrahedron, both 5$^{th}$ NN Cu2 sites in a neighboring tetrahedron will have the same probability of being vacant [Fig. 3c]. Since a vacancy can choose either of the two, there is frustration in the VO.

5$^{th}$ NN rule

The MC simulations and the 3D-ΔPDF-like analysis show that the F-forbidden peaks stem from VO. In the VO, each vacancy should find its nearest neighbour vacancies only at the 5$^{th}$ NN positions within the Cu2 network, placing the shortest inter-vacancy distance at ∼ 6.2 Å. This rule allows us to construct a $Pa\bar{3}$ supercell structure with vacancies placed at a 5$^{th}$ NN distance successfully reproducing all observed Bragg reflections (SM I.iv). We note that it is also possible to construct a twinned structure description, which reproduce the low *T* reflections, but that structure is incompatible with the observed vacancy level in the high *T* phase (SM I.iv).

The 5$^{th}$ NN rule does not define a unique VO configuration, but results in many energetically equivalent configurations. For a vacancy in a Cu$_4$ tetrahedron, there are two equivalent 5$^{th}$ NN positions in each of the 3 nearest tetrahedra [Fig. 3c]. Furthermore, a 20% vacancy level leads to a portion of the Cu$_4$ tetrahedra being fully occupied, and their mixing with the VO states brings additional complexity. All these degenerate configurations compete mutually and simultaneously freeze out when Kutinaite cools below $T_{vo}$.

*Signatures of frustrated vacancy order*

Freezing out competing degenerate configurations generally leads to a substantial residual entropy of the ordered states [5,21,22] and induces complex domain structures [23-25]. Around $T_{vo}$ Kutinaite exhibits a broad hump in the specific heat capacity $C_p$ [Fig. d, inset] and large hysteresis in the $T$-dependencies of the resistivity ρ [Fig. d] and the magnetic susceptibility χ (SM V). The hysteresis in the physical properties and the absence of a clear low-$T$ subgroup related to the high-$T$ $Fm\bar{3}m$ structure suggest a 1$^{st}$ order transition [26-28]. Furthermore, the broadness of the $C_p$ peak resembles frustration behaviors in quantum spin-ice [22] and weakly 1$^{st}$ order transitions in simulations of frustrated systems [25].

We found by integrating $C_p/T$ [Fig. 3d] that the VO reduces the total entropy by $\Delta S = \int (C_p/T)dT \approx 4.6$ J · (K·mol)$^{-1}$. In the vacancy disordered (VD) state at $T > T_{vo}$, the vacancies distribute randomly with one vacancy per Cu$_4$ tetrahedron as the sole restriction. Four equivalent vacancy positions are then possible in each tetrahedron. There are 8 Cu$_4$ tetrahedra and 4 formula units per unit cell, resulting in a total molar number of configurations of $\Omega = (4^8)^{N_A/4}$. The definition of entropy $S = k_B \ln\Omega$ yields $S_{vd} = 4R \ln 2 \approx 23.05$ J · (K · mol)$^{-1}$ as the entropy of the disordered state, where $N_A$, $k_B$, and $R$ are Avogadro's, Boltzmann's, and ideal gas constants, respectively. Subtracting the measured $\Delta S$, we obtain $S_{vo} = 18.45$ J · (K · mol)$^{-1}$ as the residual entropy of the VO state. On the other hand, at $T < T_{vo}$, because of the 5$^{th}$ NN-rule, putting a vacancy at a vertex of an arbitrary Cu$_4$ tetrahedron restricts the number of possible positions to 2 in each of the 3 NN tetrahedra and 4 in each of the remaining 4 tetrahedra in the unit cell. Considering also 4 possible positions of the first vacancy, the calculated residual entropy is then $S_{vo}^{cal} = 3.25R \ln 2 \approx 18.73$ J · (K · mol)$^{-1}$, yielding $\Delta S^{cal} \approx 4.3$ J · (K · mol)$^{-1}$. Although this calculation is oversimplified, it closely reproduces the experimental values and reproduces the crystallographic results.

Quenching a well-relaxed sample exposes the signatures of a slow formation of domain structures. We traced the intensity and the full width at half maximum (FWHM) of the F-forbidden reflections (321) and (421) and the F-allowed (222) reflection in the PXRD experiment (Fig. 2a) using single peak fitting (Fig. S5-6). Using the intensity to measure the strength of ordering [Fig. 4a], we re-confirm that the VO transition occurs on top of the stable $Fm\bar{3}m$ average structure. The (321) and (421) VO reflection dis-/re-appears upon heating/cooling around $T_{vo} \approx 300$ K, whereas the (222) reflection remains strong in the whole $T$-range.

The FWHM represents the coherence length of the scattered X-ray and is often interpreted as the amount of microstrain in the structure and/or the size of the coherently scattering domain. Fig. 4b shows a slight narrowing of all reflections in heating to 400 K, suggesting an annealing effect and release of microstrain. In a sharp contrast, the 240 K quenching greatly broadens the (421) and (321) VO reflections whereas the (222) reflection of the average structure remains virtually unaffected. If this drastic FWHM change was due to strains, the (222) reflection would also broaden. Thus, the quenching causes a strong fragmentation of the VO but leaves the average structure untouched. If the coherence scattering length measures the extent of the VO, the Debye-Scherrer estimation [Fig. 4c] gives about 40 unit-cells, fragmented down to ~1/3 of its original value. Tracing the FWHM at different cooling rates (Fig. 4c) shows that the extent of VO is directly related to the cooling rate.

The fragmentation of the frustrated VO also affects the electronic properties. The $\rho(T)$ curve [Fig. 4d] exhibits a prominent hysteresis around $T_{vo}$, where the material becomes more conductive and diamagnetic [see also Fig. 5a]. The hysteresis encircling the transition is highly asymmetric with a well-defined superheating temperature $T_{sh}$, but lacks a clear supercooling temperature $T_{sc}$ [inset of Fig. 4d]. Faster $T$-sweep rates expand the hysteresis even more to the low-$T$ side, whereas $T_{sh}$ remains constant at around 330 K. Regardless of the $T$-sweep rate, the low-$T$ VO completely dissolves at $T_{sh}$ [26].

Fig. 4e shows how the fragmentation degrades the electronic properties: faster cooling rates produce more resistive low-$T$ states and lower ratios of residual resistance (RRR). This fast-cooling effect corroborates a reduced VO coherence length in Fig. 4c. The main source of resistivity is electronic scattering at domain boundaries. The finer fragmentation caused by faster cooling rates consequently lead to more boundary states, which are unfavorable for transport. In sharp contrast, the electronic properties within each individual domain are characterized by a remarkable electronic diamagnetism.

## Diamagnetic uncompensated semimetal

Despite 20% of the Cu2 atoms missing, Kutinaite exhibits an unprecedented large metallic diamagnetism [Fig. 5a], a rare signature of non-trivial electronic properties [20]. The magnitude of the negative magnetic susceptibility, $\chi = -5.7 \times 10^{-5}$ [SI], is significantly larger than expected from Lamor's closed shell ions or ring-current (RC) mechanism (SM V). As shown in Table S4, Kutinaite's diamagnetism is comparable with the LP susceptibility of bismuth and LaV$_2$Al$_{20}$ [17,29]. The magnetic susceptibility of a nonmagnetic metal is defined by balancing the strengths of two opposite mechanisms. Pauli paramagnetic susceptibility ($\chi_P > 0$) comes from the alignment of all carriers' spins by the magnetic field and is often stronger than the magnitude of LP diamagnetism ($\chi_L < 0$), which arises from magnetic field induced cyclotron motions of carriers. Thus, metals are typically paramagnets. Owing to its dynamical origin, $|\chi_L|$ is inversely proportional to carrier effective mass $m^*$. The existence of a tiny non-trivial Fermi pocket hosting extremely light carriers underlies the rare metallic diamagnetism found in bismuth and graphite.

After excluding small Lamor's and RC contributions, we combined the results from magnetotransport and thermodynamic measurements to verify the electronic diamagnetism. The Pauli paramagnetism is proportional to $n_{\text{tot}}^{1/3} m_{\text{av}}^*$, where $n_{\text{tot}}^{1/3}$ and $m_{\text{av}}^*$ are the total number and the average effective mass of all carrier types, respectively [20]. Using $n_{\text{tot}} \approx 2.38 \times 10^{21}$ cm$^{-3}$ and $m_{\text{av}}^* \approx 17\, m_e$ ($m_e$ is the free electron mass) as estimated from Hall effect ($\rho_{yx}$) and specific heat ($C_p$), we obtain $\chi_P \approx 7.6 \times 10^{-5}$ in SI units. We then find that $\chi_L \approx -1.2 \times 10^{-4}$, being almost equal to the diamagnetism of bismuth [17,29].

A picture of an uncompensated semimetal is derived from our analyses of magnetoresistance ($\Delta\rho$) and $\rho_{yx}$ [Fig 5d] (SM III.iii), and the dynamics of the light-mass electron-like carriers in pocket 1 is the reason behind the large $\chi_\text{L}$. Pocket 1 holds a small number of electron-like carriers, $|n_1| \approx$ $1.2 \times 10^{19}$ cm$^{-3}$ [Fig 5c], but with mobility ($\mu_1$), $|\mu_1| \approx 866$ cm$^2$(Vs)$^{-1}$ [Fig 5b]. Even with the impeding scatterings at inter-grain and domain boundary inherent here, $\mu_1$ is comparable with the value of about $1400$ cm$^2$(Vs)$^{-1}$ of typical topological protected surface states in Bi$_2$Se$_3$ single crystals [30,31]. Given that $\mu_1 = e\tau_1/m_1$, the light mass $m_1$ and/or long relaxation time $\tau_1$[32], and its small size suggest that pocket 1 harbors non-trivial Dirac-like states. Consistently, $\Delta\rho$ at low $T$ also exhibits a linear dependence on magnetic fields, which is a signature of Dirac fermions following the Abrikosov model [33-35]. Using the model, we estimated that $E_\text{F} \approx 16$ meV and $v_\text{F} \approx 3.12 \times 10^5$ ms$^{-1}$ as Fermi energy and Fermi velocity of pocket 1, respectively (SM III.iv). These values fall into the same order of magnitude for other Dirac states [30,31,35]. On the other hand, the hole-like pocket 2 holds most of the total number of carriers with $n_2 \approx 200|n_1| \approx 2.37 \times 10^{21}$ cm$^{-3}$ with a lower mobility $\mu_2 \approx 100$ cm$^2$(Vs)$^{-1}$. Pocket 2 makes the major contribution to $\chi_\text{P}$ and the electronic heat capacity.

Finally, we note an upturn at $T^* \approx 60$ K in $\chi(T)$ coinciding with a change from $T$-linear to $T^2$ dependence in $\rho$. No structural change was observed between 300 K and 2 K. This feature may correspond to an electronic or magnetic transition that remains hidden.

## Discussion

Our study has discovered frustrated VO in Kutinaite. 20% vacancies on Cu2 sites not only enable the emergence of a non-trivial electronic material but also enriches its chemistry and physics with the frustrated VO. Relying on the combination of X-ray scattering methodologies, simulation, and physical properties, we deduced a simple 5$^\text{th}$ NN rule which describes the complexity surrounding the average structure, the VO, and its frustration. The concomitant occurrences of one vacancy per Cu$_4$ tetrahedron and the LP diamagnetism are highly reminiscent of the case of Bismuth, in which a slight deformation from cubic structure gives rise to a most intriguing electronic state [2]. Somehow the crystal structure stabilizes the defect Cu tetrahedra, and unlike the fully occupied structure with conventional properties, the defect structure attains non-trivial electronic states. In some sense one may view the vacancy as a "chemical species". Full understanding of the vacancy ordering and the unprecedented metallicity must await future work on chemical bonding to explain why vacancies repel each other until they are 6.2 Å away, and how non-trivial electronic states arise from the missing Cu2 vertices.

## Methods

### Synthesis

Polycrystalline Kutinaite was synthesized by a solid-state chemical reaction between high purity copper, silver, and arsenic at 1363 K for 10 hours in an evacuated silica tube, followed by a 3-day annealing at 623 K. The starting composition was chosen at Ag:Cu:As = 6:14:7 following a previous report[13]. The obtained sample was ground and sieved down to 200 μm, then densified using sparked-plasma-sintering method. The resulting Kutinaite samples have a density higher than 95% of the value estimated from SCXRD.

### Single crystal X-ray diffraction

Single crystals (approx. 50x50x50 μm$^3$) were selected from the ground sintered ingot and used for SCXRD measurements with 50 keV (λ = 0.2482 Å) X-ray radiation at BL02B1 beamline, SPring-8. The diffractometer was equipped with a quarter-χ goniometer and a Pilatus 3 X 1M CdTe detector. Three 180° ω scans were performed at χ = 0°, 20°, and 45° with a detector distance of 130.0 mm and frame width of 0.1 °/frame. Attenuation was estimated based on the maximum flux from a screening experiment[36]. Structural datasets were collected at six different temperatures while cooling (350 K, 310 K, 290 K, 250 K, 100 K, 25 K) and two temperatures upon subsequent heating (100 K and 300 K). A Cu200μm attenuator was used for the 350 K, 310 K and 290 K datasets and Ni400μm for the 100 K dataset, no attenuation was used at 25 K or for the datasets on heating. An exposure time of 0.1 s/frame was used, except for the 350 K dataset with 0.2 s/frame where also the first ω scan at χ = 0° was excluded due to temperature fluctuations. Images were converted (code available at: https://github.com/LennardKrause/) and integrated using SAINT[37,38] with a resolution of (sinθ/λ)$_{max}$ = 0.83 Å$^{-1}$ (d$_{min}$ = 0.6 Å) for all datasets. Processing and multi-scan absorption correction (μ = 4.286 mm$^-$

1, μr = 0.12) was performed in SADABS[39]. The structure was solved and refined in OLEX2[40] using SHELXT[41] and SHELXL[42], respectively. Atomic positions and anisotropic atomic displacement parameters (ADPs) were refined for all atoms as allowed by the site symmetry to describe harmonic vibrations, and the occupancy of Cu2 was refined freely at all temperatures. The structure refinement quality parameters are $R_{int} < 0.075, R(F^2) < 0.02, wR(F^2) < 0.05$, GOF $\approx 1.0$, and residual density $|\Delta\rho| \leq 1 e\text{Å}^{-3}$ at all temperatures from 25 K to 350 K. Further crystallographic and structural information, as well as a discussion of the modelling of anharmonic ADPs can be found in the Supplementary Section I.i.

For analysing the disorder correlations, a method related to the three-dimensional difference pair-distribution function (3D-ΔPDF) was developed. This method uses the pair-distribution function arising from the difference between the PDF from the total scattering and the PDF from the average structure[15]. The common procedure for obtaining the 3D-ΔPDF is by removing the Bragg peaks from the total scattering pattern followed by Fourier transforming the remaining diffuse scattering[16]. To understand the correlations giving rise to a P-centered cell, the Bragg peaks conforming to the F-centered cell were removed from the structure factor file. That is, a difference Fourier map (DFM) was constructed by Fourier transforming only the peaks fulfilling at least one of $h + k \neq 2n$, $h + l \neq 2n'$ or $k + l \neq 2n''$.

### Powder X-ray diffraction

The synchrotron PXRD experiment was performed at the RIKEN Materials Science I beamline BL44B2 at the SPring-8 Synchrotron Facility in Hyogo, Japan across two separate beamitimes.[43] The PXRD data were collected using the OHGI detector.[44] A powder sample of the polycrystalline Kutinaite were packed in a 0.2 mm glass capillary and the PXRD data were collected with an exposure of 180 s per temperature in both experiments. The temperature was controlled using a Nitrogen cold gun.

Temperature dependent PXRD experiments were performed by cooling to 240 K and then increasing the temperature in steps of 20 K until the temperature of 400 K was reached. After the first heating cycle, the sample was cooled to 300 K and 240 K. The cooling should be regarded as quenching as the heat is removed the temperature drops immediately due to the small size of the sample. For these experiments the beamline was operated using a photon energy of 21.80 keV (λ= 0.56974 Å).

Secondly, the effect of cooling rate was investigated by heating to 360 K and the cooling to 300 K using temperature rates of: 2, 5, 10, 25, 50, 100 Kmin$^{-1}$. For these experiments the beamline was operated using a photon energy of 21.40 keV (λ= 0.58000 Å).

All Rietveld refinements were performed in Topas V7.[45] The average structure was modelled based on the structure obtained from the single crystal X-ray experiments. The instrumental peak shape was determined from a NIST Si 640d line broadening standard along with sample displacement and X-ray wavelength. The background was described by a Lorentzian peak centred at 7.9° together with a gaussian peak at 25.4° and a 5$^{th}$ degree Chebyshev polynomial. The structural parameters refined were the scale factor, unit cell, atomic position, ADPs and Lorentzian microstrain. All the obtained structural parameters are summarized in Tab. S3.

Single peak fitting was performed in Python using the lmfit minimizing algorithm. The peak was fitted using a pseudo-Voight peak on a linear background. A single Lorentzian parameter was refined for each peak. The remaining peak shape parameters were fixed to the values obtained from the Si NIST reference. The domain size was estimated using the Scherrer formular.

A vacancy ordering model were obtained using a 5x5x5 supercell of the average structure in P1 symmetry. Ordering of the vacancies were performed using a simple Monte Carlo algorithm written in Python. To evaluate the vacancy models a simple Rietveld refinement of the 240 K dataset was performed in Topas v7 using the supercells in which only the scale factor, background and unit cell were refined. The remaining structural parameters were fixed to the values obtained from the average structure.

### Electric transport

Electrical transport properties of Kutinaite were measured using standard lock-in amp technique with the help of Quantum Design (QD) Electrical Transport Option (ETO) and Physical Properties Measurement System (PPMS). The temperature and magnetic field ranges of 2 K – 300 K and ±9 T, respectively. High-density sintered Kutinaite ingots were cut and polished into rectangular box samples. The lengths of the samples vary between 4 and 2 mm, and width and thickness are smaller than 2.5 mm and 1 mm, respectively. To measure both Hall effects and magnetoresistance, six electrodes were made on each sample by DuPont silver paste. We measured the temperature and magnetic field dependencies of the transverse ($\rho_{yx}$ or Hall) and longitudinal ($\rho_{xx}$) resistivities with an AC current at frequencies less than 50 Hz The magnetotransport data reported hereafter were symmetrized between positive and negative magnetic fields to eliminate the unwanted contribution from geometric misalignments of electrodes, i.e. $\rho_{xx}(H) = 0.5 \times \left(\rho_{xx}^{meas}(H) + \rho_{xx}^{meas}(-H)\right)$ and $\rho_{yx}(H) = 0.5 \times \left(\rho_{yx}^{meas}(H) - \rho_{yx}^{meas}(-H)\right)$, with $\rho_{ij}^{meas}$ are experimentally measured values. To estimate the effect of cooling rate on the RRR, we monitored and averaged the steady-state resistivity in 15 minutes after cooling (warming) from (to) 5 K (400 K) at various ramping rates.

### Magnetic properties

We investigated the magnetism of Kutinaite by Quantum Design (QD) Magnetic Properties Measurement System 3 (MPMS-3) in the temperature range from 2 K to 400 K. Magnetizations were measured by both DC and Vibrating Sample Magnetometry (VSM) techniques at various temperatures and magnetic fields. The difference between DC and VMS values is less than 1 %.

### Heat capacity

The specific heat capacity of Kutinaite was measured by relaxation-time method using the QD-PPMS Heat Capacity option. A sample was put on the platform, and the thermal contact was maintained by Apiezon grease N or H. The heat capacity of the grease was measured and used as addenda.

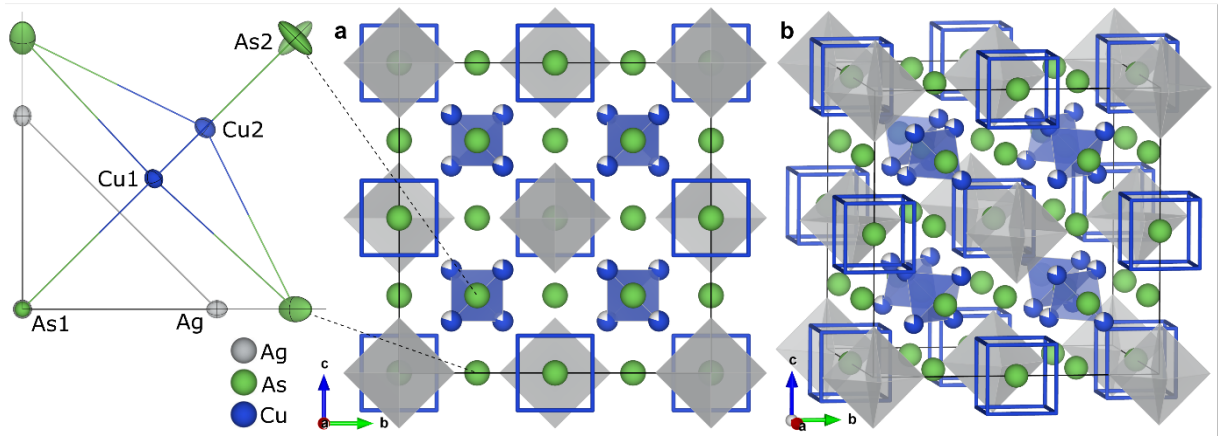

Fig. 1. Unit cell of the average structure of Kutinaite solved by single crystal X-ray diffraction at 25K (space group Fm-3m). (a) structure viewed down the a-axis and (b) slightly askew. Polyhedrons and bonds are visualized to guide the eye. Ag is placed in the vertices of the grey octahedrons, eight-fold coordinated As1 (green) is placed in the center of the Cu1 (blue) cubes, while As2 (green) is shown as independent spheres. The partial occupancy of Cu2, placed in the vertices of the blue tetrahedrons, in the average structure is shown by partially filled spheres. The insert in (a) shows the structure (a < 0.5, b < 0.25, c < 0.25) with anisotropic vibrational parameters (50% probability ellipsoids).

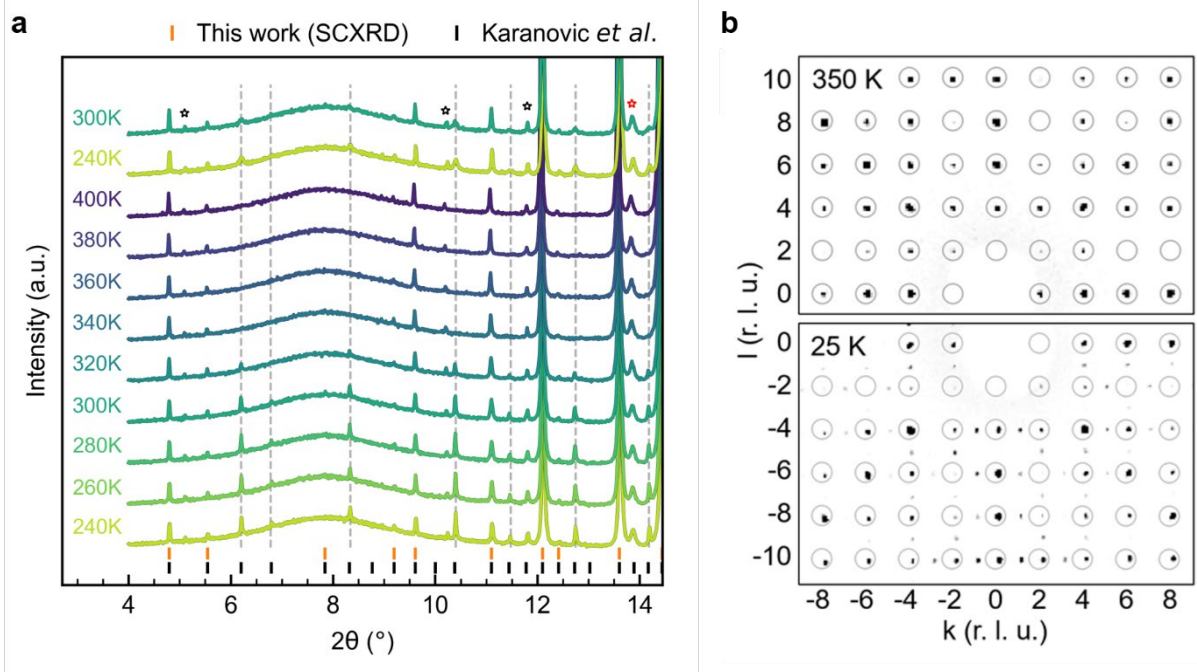

Fig. 2. Temperature dependencies of PXRD and SCXRD data. (a) PXRD from 240 to 400 $K$. The orange ticks correspond to the reflections expected for the $Fm\overline{3}m$ structure whereas the black ticks correspond to the $Pm\overline{3}m$ structure. The black and red stars highlight the As4O6 and Ag impurities. The grey dashed line is a guide to the eye highlighting the additional F-forbidden reflections. (b) Reconstructed precession images of SCXRD in the $0kl$ plane at $350\ K$ and $25\ K$. Indexes are given in reciprocal lattice units (r. l. u.). Grey circles indicate the peaks indexed by the $Fm\overline{3}m$ unit cell.

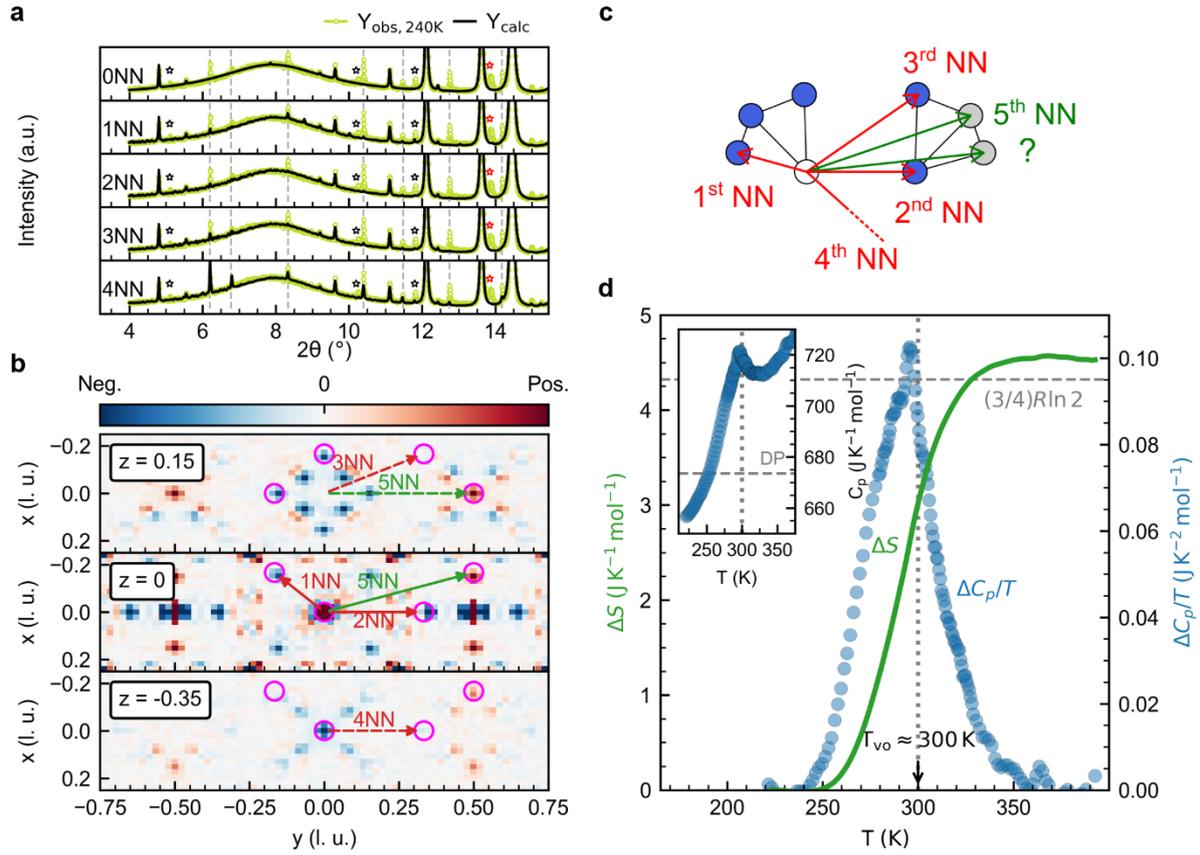

Fig. 3. Competing vacancy orders. (a) Rietveld refinement of PXRD data measured at 240 K using MC models 1NN to 4NN. (b) The z = 0, 0.15, and -0.35 planes of the 3D-DPDF-like map with vectors to the nearest neighbors using the same origin as in [Fig. 3c]. The magenta circles mark the location of the Cu2-atoms relative to the chosen origin. The dashed lines remind that these vectors have their origin in the z = 0 plane. (c) Schematics showing the frustrated ordering of vacancies in neighboring tetrahedra. (d) Specific heat capacity $C_p$ around $T_{vo}$ dominated by broad transitional hump (insert). Unavoidable microstrain introduced by the synthesis process raises the total $C_p$ above the Dulong-Petit limit (DP). A linear background was subtracted from the data to obtain $C_p/T$ (blue circles) and the change of entropy $\Delta S$ (bold green line) around $T_{vo}$. The black dashed line shows the value of $\Delta S^{cal} = (3/4)R \ln 2$ estimated by counting all possible vacancy configurations of the disordered state at $T > T_{vo}$ and competing VOs at $T < T_{vo}$ (see text).

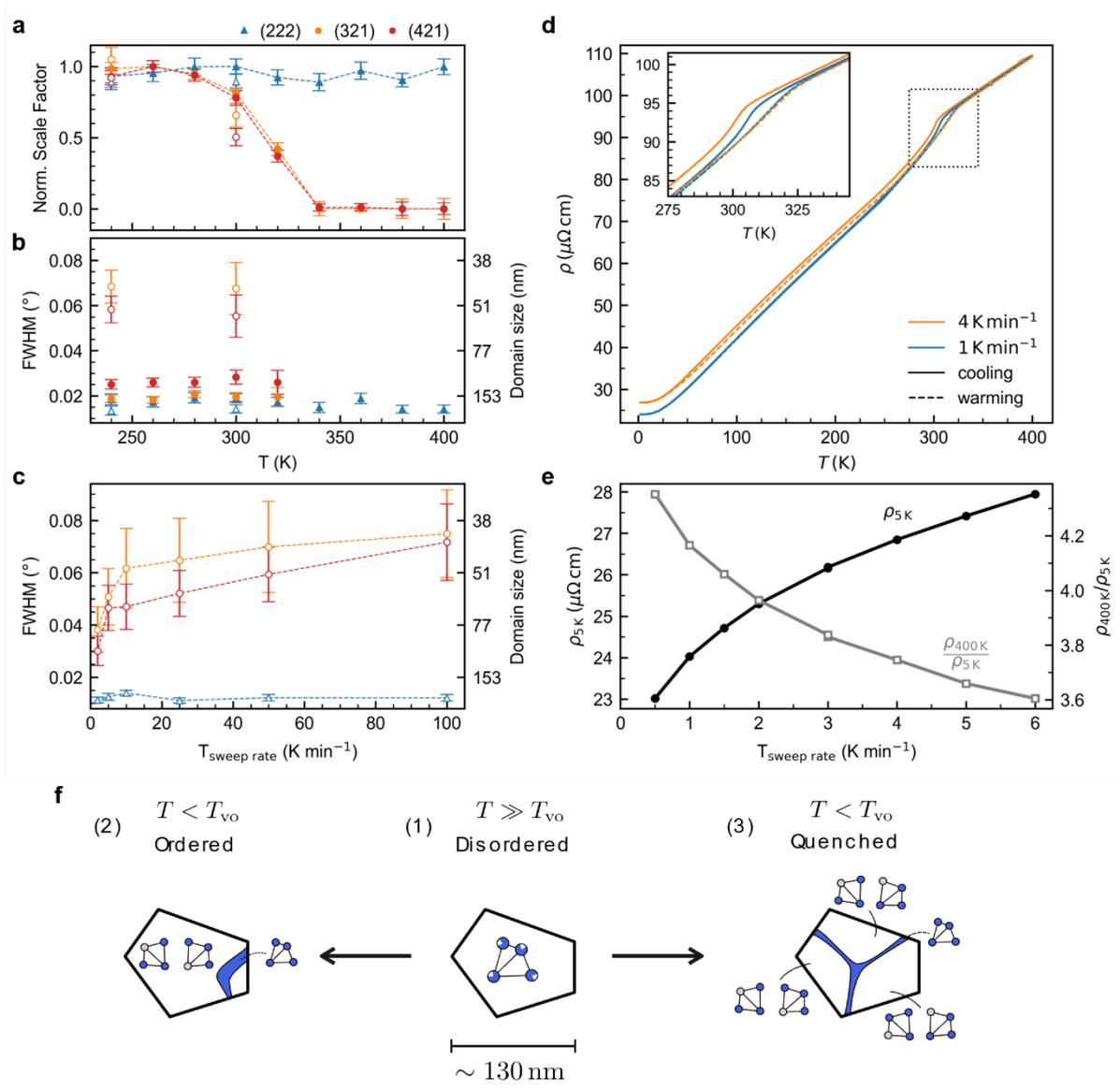

Fig. 4. Thermal cycling effects of disorder. (a) Normalized scale factor and (b) FWHM and coherence length resulting from single peak fitting analyses of the (222) (average structure), (321) and (421) (VO) reflections. The empty symbols denote values obtained upon cooling. (c) FWHM of the (222), (321) and (421) reflections at 300 K upon cooling with varying sweep-rate. (d) Resistivity $\rho(T)$ of Kutinaite measured from 2 K to 400 K at $T$-sweep rates of 1 K/minute and 4 K/minute. The inset shows the hysteresis around $T_{vo}$. (e) Sweep-rate dependence of residual resistivity at $T = 5$ K and ratio-of-resistivity $\rho(400\,\text{K})/\rho(5\,\text{K})$. (f) Simplified cartoons illustrating various structures of a Kutinaite grain suggested by the diffraction experiment in (a) and (b). Neither the size nor the average crystal structure of the grain changes throughout the whole process. (2) presents the well-relaxed ordering state at the beginning of the experiment with rather large coherence length. (1) Warming the sample above $T_{vo}$ destroys this order and vacancies are distributed randomly. (3) Quenching from high $T$ freezes the competing degenerated VOs and results in a shorter range of coherent ordering.

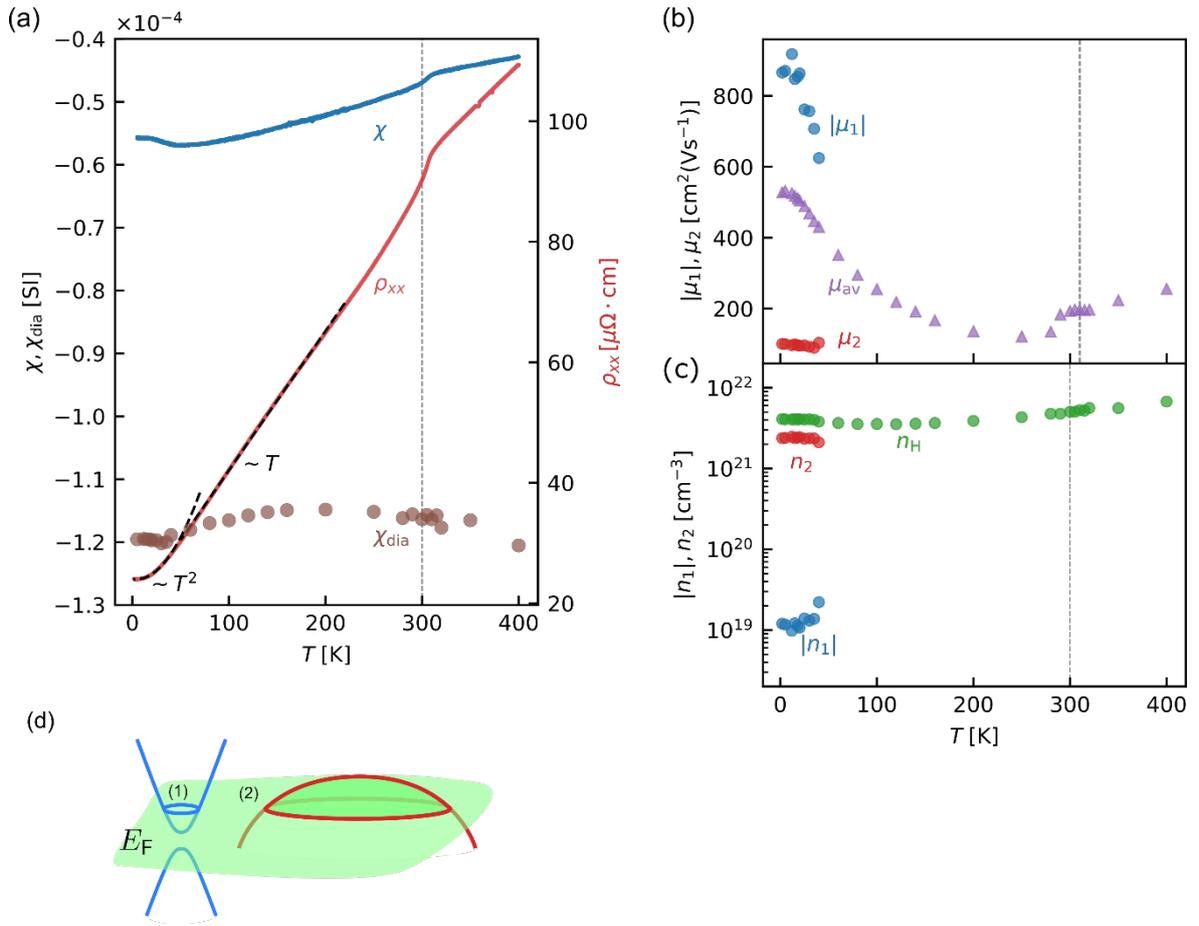

Fig. 5: Non-trivial electronic properties of Kutinaite. (a) Coexistence of metallicity and diamagnetism in Kutinaite. The measured magnetic susceptibility $\chi$ is negative, and the resistivity $\rho$ monotonically increases with increasing $T$. $\chi_L$ is diamagnetic susceptibility after subtracting other magnetic contributions. (b) and (c) show the mobilites and numbers of electron-like (1) and hole-like (2) carrier types resulting from our analyses of magnetoresistance and Hall effect. The number of hole-like carriers is two orders larger than that of electron-like carriers. On the other hand, electron-like carriers have much higher mobility. The VO displays itself as a clear kink at $T_{vo}$ in all physical properties. $\mu_i$ and $n_i$ ($i = 1, 2$) are estimated from 2-carrier-type analysis; $\mu_{av}$ and $n_H$ are estimated from the low-field expansion of the same model (SM III.iii). (d) Schematic illustration of the suggested electronic structure of Kutinaite as gathered from $C_p$, Hall effects, and magnetoresistances. The hole-like carrier type 2 occupying the large Fermi pocket with a heavy effective mass and gives the dominant contribution to $C_p$ and $\chi_P$.

# Frustrated vacancy ordering creates novel quantum properties in Kutinaite, $Ag_6Cu_{14.4}As_7$

Kim-Khuong Huynh*[a], Rasmus Baden Stubkjær[a], Ventrapati Pavankumar[b], Emilie Skytte Vosegaard[a], Karl Omer Rimon Juul[a], Kasper Rasmussen Borup[a], Yong P. Chen[c], and Bo Brummerstedt Iversen*[a]

[a]Center for Integrated Materials Research, Department of Chemistry and iNANO, Aarhus University, Langelandsgade 140, 8000 Aarhus (Denmark)

[b]Department of Chemistry, SRM University Amravati-AP, Mangalagiri 522240 AndhraPradesh, India

[c]Center for Integrated Materials Research, Department of Physics and iNANO, Aarhus University, Langelandsgade 140, 8000 Aarhus (Denmark)

*Corresponding authors: hkkhuong@chem.au.dk, bo@chem.au.dk

## 1 Single crystal X-ray diffraction

### 1.1 Experimental details

**Tab. S1. Crystallographic information for the *Fm-3m* structures.** *Datasets measured upon heating.

| Chemical formula | $Ag_6As_7Cu_{14.371}$ | $Ag_6As_7Cu_{14.417}$ | $Ag_6As_7Cu_{14.392}$ | $Ag_6As_7Cu_{14.39}$ | $Ag_6As_7Cu_{14.385}$ | $Ag_6As_7Cu_{14.349}$ | $Ag_6As_7Cu_{14.349}$ | $Ag_6As_7Cu_{14.399}$ |
|---|---|---|---|---|---|---|---|---|
| $M_r$ | 2084.73 | 2087.75 | 2086.16 | 2086.00 | 2085.68 | 2083.46 | 2083.46 | 2086.64 |
| Crystal system, space group | Cubic, *Fm-3m* | Cubic, *Fm-3m* | Cubic, *Fm-3m* | Cubic, *Fm-3m* | Cubic, *Fm-3m* | Cubic, *Fm-3m* | Cubic, *Fm-3m* | Cubic, *Fm-3m* |
| Temperature (K) | 25 | 100 | 100* | 250 | 290 | 300* | 310 | 350 |
| $a$ (Å) | 11.7350(2) | 11.7475(2) | 11.7461(3) | 11.7690(5) | 11.7867(4) | 11.7829(9) | 11.7871(9) | 11.7942(8) |
| $V$ (Å$^3$) | 1616.03(10) | 1621.20(10) | 1620.62(12) | 1630.12(19) | 1637.48(15) | 1635.9(4) | 1637.6(4) | 1640.6(3) |
| $Z$ | 4 | 4 | 4 | 4 | 4 | 4 | 4 | 4 |

| $T_{min}$, $T_{max}$ | 0.634, 0.836 | 0.634, 0.836 | 0.628, 0.836 | 0.670, 0.836 | 0.692, 0.836 | 0.599, 0.836 | 0.620, 0.789 | 0.540, 0.748 |
|---|---|---|---|---|---|---|---|---|
| No. of measured, independent and observed [$I > 2\sigma(I)$] reflections | 19376, 241, 234 | 19780, 248, 237 | 19567, 248, 238 | 19693, 241, 228 | 19990, 248, 237 | 18715, 241, 230 | 19012, 242, 230 | 11939, 240, 227 |
| $R_{int}$ | 0.038 | 0.042 | 0.040 | 0.053 | 0.051 | 0.049 | 0.069 | 0.074 |
| $R[F^2 > 2\sigma(F^2)]$, $wR(F^2)$, $S$ | 0.008, 0.017, 1.15 | 0.009, 0.020, 1.11 | 0.009, 0.020, 1.11 | 0.012, 0.027, 1.13 | 0.011, 0.026, 1.19 | 0.015, 0.038, 1.18 | 0.014, 0.036, 1.12 | 0.019, 0.050, 1.16 |
| $\Delta\rho_{max}$, $\Delta\rho_{min}$ (e Å$^{-3}$) | 0.53, −1.02 | 0.47, −0.86 | 0.47, −0.86 | 0.48, −0.83 | 0.41, −0.79 | 0.57, −0.89 | 1.07, −0.80 | 0.86, −0.95 |

## 1.2 Anharmonic structural refinement

Anharmonic vibrations were modelled by refining 4$^{th}$ order Gram-Charlier (GC) coefficients, $\delta^{jklm}$,[1] on As2 in the program XD2016[2] for the 25 K dataset solved in the *Fm-3m* space group up to a resolution of (sinθ/λ)$_{max}$ = 1.20 Å$^{-1}$ (d$_{min}$ = 0.42 Å) including 9744 reflections. All symmetry allowed coefficients ($\delta^{1111} = \delta^{2222}, \delta^{3333}, \delta^{1112} = \delta^{1222}, \delta^{1122}, \delta^{1133} = \delta^{2233}, \delta^{1233}$) refines to values significant within three standard deviations. The nuclear probability density function (nPDF) was evaluated to assess the reliability of the model, and a low negative probability density of -2.6% was found. The anharmonic refinement improves the model significantly by improving the quality parameters ($R(F^2) = 3.51\% \rightarrow 2.80\%, \Delta\rho_{min} = -8.8\ e/Å^3 \rightarrow -1.8\ e/Å^3, \Delta\rho_{max} = 4.6\ e/Å^3 \rightarrow 1.6\ e/Å^3$) and reducing the systematic residuals of alternating sign around the As2 site (Fig. S2) GC coefficients are very sensitive to high order data, so the not-negligible negative nPDF is attributed the truncated resolution range available in the dataset.

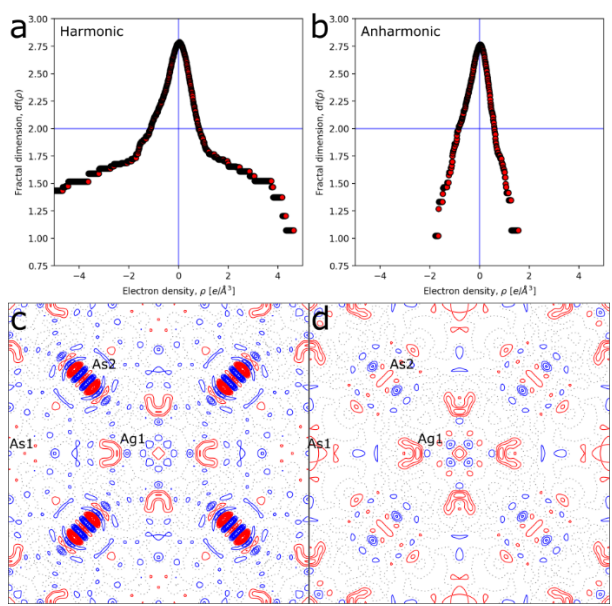

**Fig. S1. Fractal dimension plots and residual density maps.** Fractal dimension plots for a) the harmonic model and b) the anharmonic model at 25 K. Residual density maps in the (001) plane for c) the harmonic model and d) the anharmonic model with positive and negative contours in blue and red, respectively, shown with 0.5 e/Å$^3$ intervals to the full resolution of 1.2 Å$^{-1}$.

### 1.3 Cu2 occupancy

The occupancy of the Cu2 site was refined freely at all temperatures and gave consistent results of approx. 80% occupancy (Fig. S2) in the *Fm-3m* structure solution.

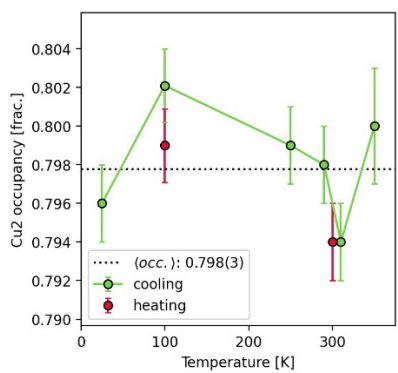

**Fig. S2. Cu2 atomic occupancy as a function of temperature.** The dotted line indicates the mean occupancy of 0.798(3). The occupancy seems to vary slightly with temperature, but most data points are within one standard deviation of the mean and all are well within 3σ (indicated by the plot limits of the y-axis). Data points collected on heating are marked with red.

## 1.4 Discussion of space group and supercell

The *Fm-3m* space group reported in the main text does not describe all observed reflections. The quality parameters reported in Tab. S1 are obtained with outlier rejection of the F-forbidden peaks. Tab. S2 shows quality parameters for the structural solution in the space group *Fm-3m* at 25 K including all reflections (also the F-forbidden). The fit is only slightly worse than the reduced dataset used in Tab. S1, due to the low intensity of the F-forbidden peaks. A twinned structure obeying the 5NN-rule can be solved in the space group *Pa-3,* which uses all reflections. By the 5NN-rule the choice of the first vacancy gives two options in the nearest neighbor tetrahedron, but after this, the restriction of symmetry locks the rest of the positions. This results in two configurations that can be related by a 90° rotation around any of the three cell axis, and can therefore be assigned as a twin in the single crystal structure solution. Quality parameters for the structure solution can be seen in Tab. S2. The twinning fraction refines to 0.500(2) with a 33% occupancy of Cu on the vacancy site, and 100% occupancy of the remaining three positions in the tetrahedron. Due to the lower symmetry the structure has more degrees of freedom and a relaxation of the structure surrounding the vacant site can be observed at the As2 and Cu2 positions (retaining the *Fm-3m* labelling) (Fig. S3). The *Pa-3* residual density map also exhibits features of anharmonic motion. The solution however gives some inconsistencies with the Fm-3m solution: the nominal composition is $Ag_6Cu_{14.7}As_7$ which is closer to the charge balanced $Cu_{15}$ composition than the previous $Ag_6Cu_{14.4}As_7$, but it means the 1/3 tetrahedra are fully occupied which is not consistent with the 1/5 in the high T phase.

**Tab. S2. Quality parameters for the *Fm-3m* and *Pa-3* structures including all reflections.**

| Chemical formula | $Ag_6As_7Cu_{14.437}$ | $Ag_6As_7Cu_{14.676}$ |
|---|---|---|
| $M_r$ | 2089.02 | 2104.11 |
| Crystal system, space group | Cubic, *Fm-3m* | Cubic, *Pa-3* |
| Temperature (K) | 25 | 25 |
| $a$ (Å) | 11.7350 (2) | 11.7350 (2) |
| $V$ (Å³) | 1616.03 (10) | 1616.03 (10) |
| $Z$ | 4 | 4 |
| $T_{min}$, $T_{max}$ | 0.5921, 0.7438 | 0.634, 0.836 |
| No. of measured, independent and observed [$I > 2\sigma(I)$] reflections | 19578, 240, 240 | 79645, 1386, 1334 |
| $R_{int}$ | 0.0425 | 0.0513 |
| $R[F^2 > 2\sigma(F^2)]$, $wR(F^2)$, $S$ | 0.0154, 0.0493, 1.210 | 0.0236, 0.0831, 1.169 |
| $\Delta\rho_{max}$, $\Delta\rho_{min}$ (e Å$^{-3}$) | 1.056, -1.284 | 2.608, -2.860 |

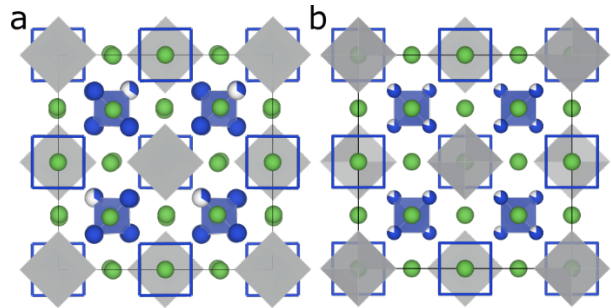

**Fig. S3. Structure solution in different space groups.** a) *Pa-3* and b) *Fm-3m* space groups using the same structure factor file with all reflections included.

## 2 Powder X-ray diffraction

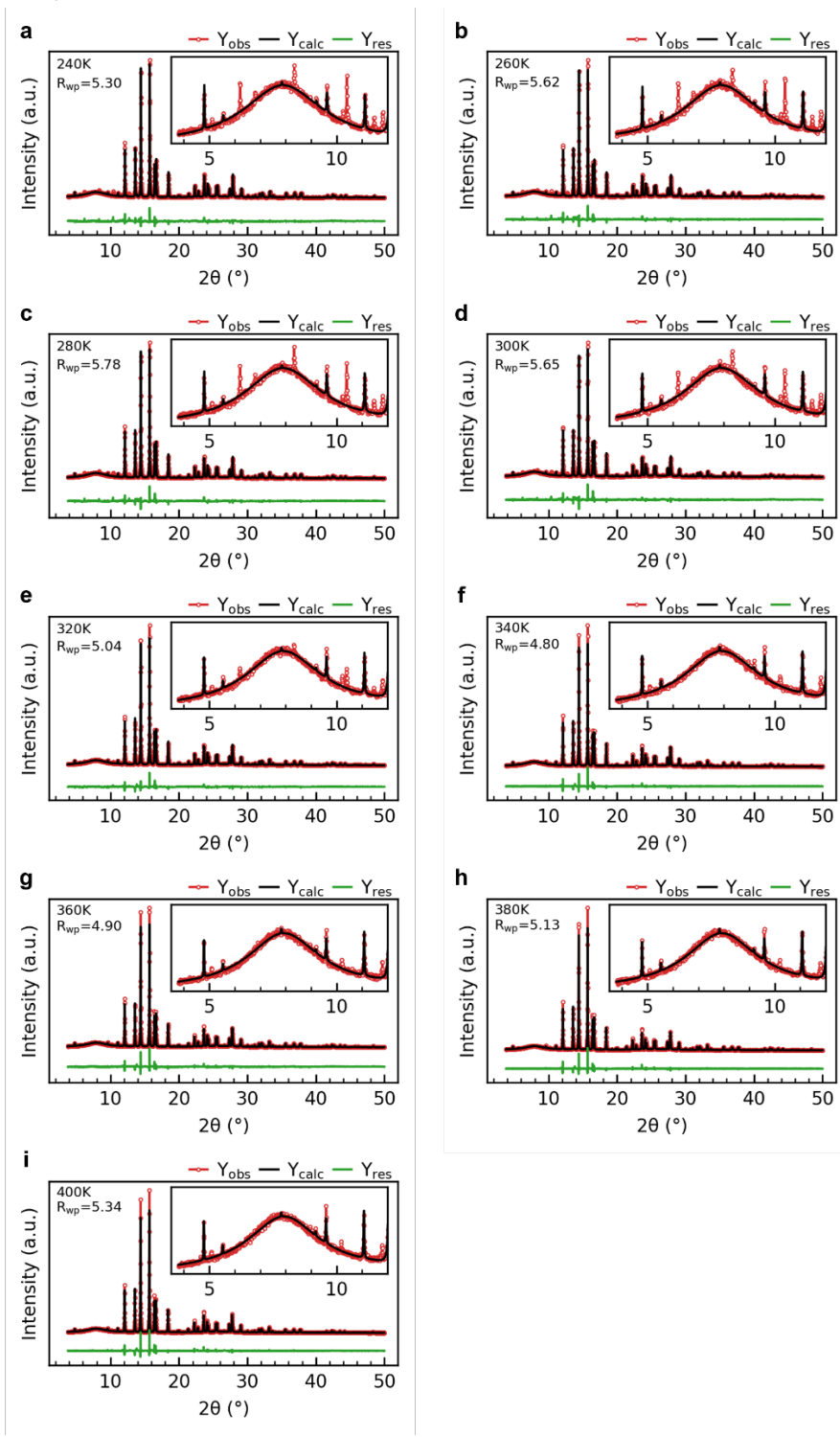

**Fig. S4. Rietveld refinement results at various temperatures.** The refinements are performed using the *Fm-3m* structure. a) 240 K, b) 260 K, c) 280 K, d) 300 K, e) 320 K, f) 340 K, g) 360 K, h) 380 K, and i) 400 K.

**Tab S3. Rietveld refinement result of the temperature dependent PXRD using the *Fm-3m* structure.**

| T [K] | 240 | 260 | 280 | 300 | 320 | 340 | 360 | 380 | 400 | 240 | 300 |
|---|---|---|---|---|---|---|---|---|---|---|---|
| $u_{11, Ag1}$ [Å²] | 0.0055 ±0.0004 | 0.0058 ±0.0004 | 0.0056 ±0.0005 | 0.0058 ±0.0005 | 0.0081 ±0.0005 | 0.0099 ±0.0005 | 0.0116 ±0.0005 | 0.0130 ±0.0005 | 0.0140 ±0.0006 | 0.0070 ±0.0004 | 0.0079 ±0.0005 |
| $u_{22, Ag1}$ [Å²] | 0.0225 ±0.0006 | 0.0284 ±0.0007 | 0.0332 ±0.0008 | 0.0378 ±0.0008 | 0.0376 ±0.0007 | 0.0396 ±0.0007 | 0.0397 ±0.0007 | 0.0421 ±0.0008 | 0.0440 ±0.0008 | 0.0227 ±0.0007 | 0.0367 ±0.0008 |
| $u_{11, As1}$ [Å²] | 0.0049 ±0.0006 | 0.0050 ±0.0007 | 0.0047 ±0.0007 | 0.0067 ±0.0007 | 0.0072 ±0.0007 | 0.0089 ±0.0007 | 0.0098 ±0.0007 | 0.0102 ±0.0007 | 0.0102 ±0.0008 | 0.0045 ±0.0006 | 0.0074 ±0.0007 |
| $u_{11, As2}$ [Å²] | 0.025 ±0.001 | 0.026 ±0.001 | 0.027 ±0.001 | 0.029 ±0.001 | 0.033 ±0.001 | 0.032 ±0.001 | 0.033 ±0.001 | 0.036 ±0.001 | 0.037 ±0.001 | 0.024 ±0.001 | 0.030 ±0.001 |
| $u_{22, As2}$ [Å²] | 0.019 ±0.001 | 0.020 ±0.001 | 0.022 ±0.001 | 0.023 ±0.001 | 0.020 ±0.001 | 0.021 ±0.001 | 0.022 ±0.001 | 0.023 ±0.001 | 0.024 ±0.001 | 0.020 ±0.001 | 0.024 ±0.001 |
| $u_{13, As2}$ [Å²] | 0.016 ±0.002 | 0.017 ±0.002 | 0.018 ±0.002 | 0.020 ±0.002 | 0.016 ±0.002 | 0.016 ±0.002 | 0.016 ±0.002 | 0.017 ±0.002 | 0.018 ±0.002 | 0.017 ±0.002 | 0.020 ±0.002 |
| $u_{11, Cu1}$ [Å²] | 0.0125 ±0.0003 | 0.0135 ±0.0003 | 0.0145 ±0.0003 | 0.0154 ±0.0004 | 0.0164 ±0.0003 | 0.0178 ±0.0003 | 0.0186 ±0.0003 | 0.0208 ±0.0004 | 0.0221 ±0.0004 | 0.0125 ±0.0003 | 0.0158 ±0.0003 |
| $u_{12, Cu1}$ [Å²] | -0.0009 ±0.0004 | -0.0020 ±0.0005 | -0.0026 ±0.0005 | -0.0045 ±0.0005 | -0.0036 ±0.0005 | -0.0040 ±0.0005 | -0.0038 ±0.0005 | -0.0033 ±0.0005 | -0.0031 ±0.0006 | -0.00180 ±0.0004 | -0.0038 ±0.0005 |
| $u_{11, Cu2}$ [Å²] | 0.0110 ±0.0003 | 0.0115 ±0.0003 | 0.0123 ±0.0003 | 0.0133 ±0.0004 | 0.0140 ±0.0003 | 0.0153 ±0.0003 | 0.0160 ±0.0003 | 0.0172 ±0.0004 | 0.0177 ±0.0004 | 0.0107 ±0.0003 | 0.0132 ±0.0003 |
| $u_{12, Cu2}$ [Å²] | 0.0024 ±0.0004 | 0.0024 ±0.0005 | 0.0023 ±0.0005 | 0.0026 ±0.0005 | 0.0036 ±0.0005 | 0.0042 ±0.0005 | 0.0042 ±0.0005 | 0.0056 ±0.0005 | 0.0056 ±0.0005 | 0.0028 ±0.0004 | 0.0033 ±0.0005 |
| Strain* | 0.291 ±0.001 | 0.291 ±0.001 | 0.292 ±0.001 | 0.294 ±0.001 | 0.283 ±0.001 | 0.2525 ±0.0009 | 0.2416 ±0.0009 | 0.2298 ±0.0009 | 0.2189 ±0.0009 | 0.2242 ±0.0009 | 0.222 ±0.001 |
| a [Å] | 11.76566 ±0.00004 | 11.77023 ±0.00005 | 11.77530 ±0.00005 | 11.78032 ±0.00005 | 11.78685 ±0.00004 | 11.79339 ±0.00004 | 11.79905 ±0.00004 | 11.80434 ±0.00004 | 11.80994 ±0.00004 | 11.76730 ±0.00004 | 11.78243 ±0.00004 |
| z (Ag1) | 0.17936 ±0.00009 | 0.1795 ±0.0001 | 0.1793 ±0.0001 | 0.1794 ±0.0001 | 0.1797 ±0.0001 | 0.1800 ±0.0001 | 0.1802 ±0.0001 | 0.1802 ±0.0001 | 0.18000 ±0.0001 | 0.1794 ±0.0001 | 0.1793 ±0.0001 |
| x (Cu1) | 0.37886 ±0.00007 | 0.37905 ±0.00007 | 0.37905 ±0.00008 | 0.37894 ±0.00007 | 0.37931 ±0.00007 | 0.37928 ±0.00007 | 0.37933 ±0.00007 | 0.37936 ±0.00007 | 0.379381 ±0.00008 | 0.37900 ±0.00007 | 0.37946 ±0.00007 |
| x (Cu2) | 0.16713 ±0.00009 | 0.16749 ±0.00009 | 0.1678 ±0.0001 | 0.1679 ±0.0001 | 0.16770 ±0.00009 | 0.16751 ±0.00008 | 0.16748 ±0.00008 | 0.16749 ±0.00009 | 0.167463 ±0.00009 | 0.16749 ±0.00009 | 0.16787 ±0.00009 |
| $R_{wp}$ [%] | 5.30 | 5.62 | 5.78 | 5.65 | 5.04 | 4.80 | 4.90 | 5.13 | 5.34 | 5.82 | 5.79 |

*Strain parameter as obtained from Topas. To obtain the value in percent a factor of $2\pi/180$ should be applied.

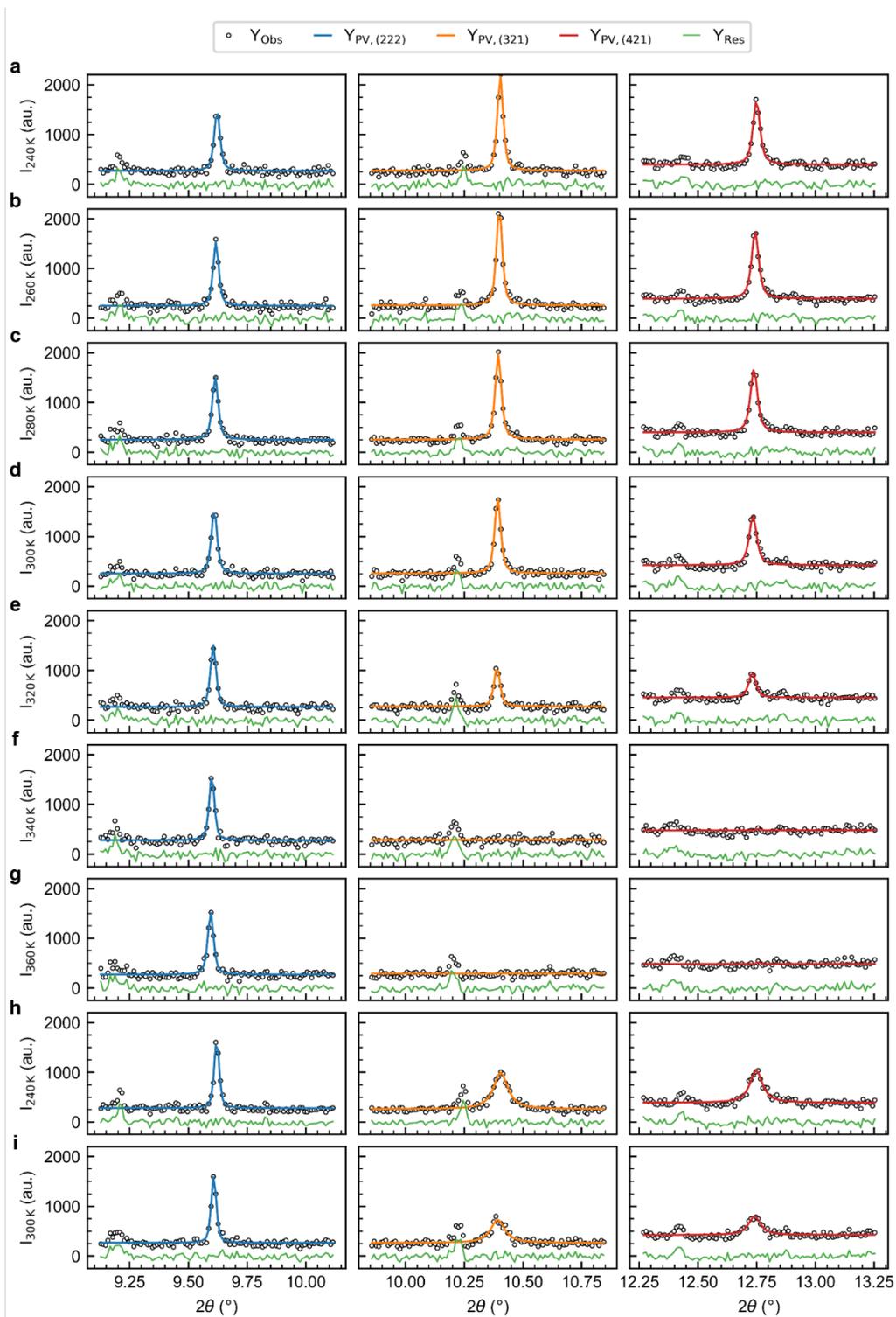

**Fig. S5. Single peak fitting at various temperatures.** Single peak fitting of the (222) (blue), (321) (orange), and (421) (red) collected at a) 240 K, b) 260 K, c) 280 K, d) 300 K, e) 320 K, f) 340 K, g) 360 K, h) 240 K (after quenching), and i) 300 K (after quenching).

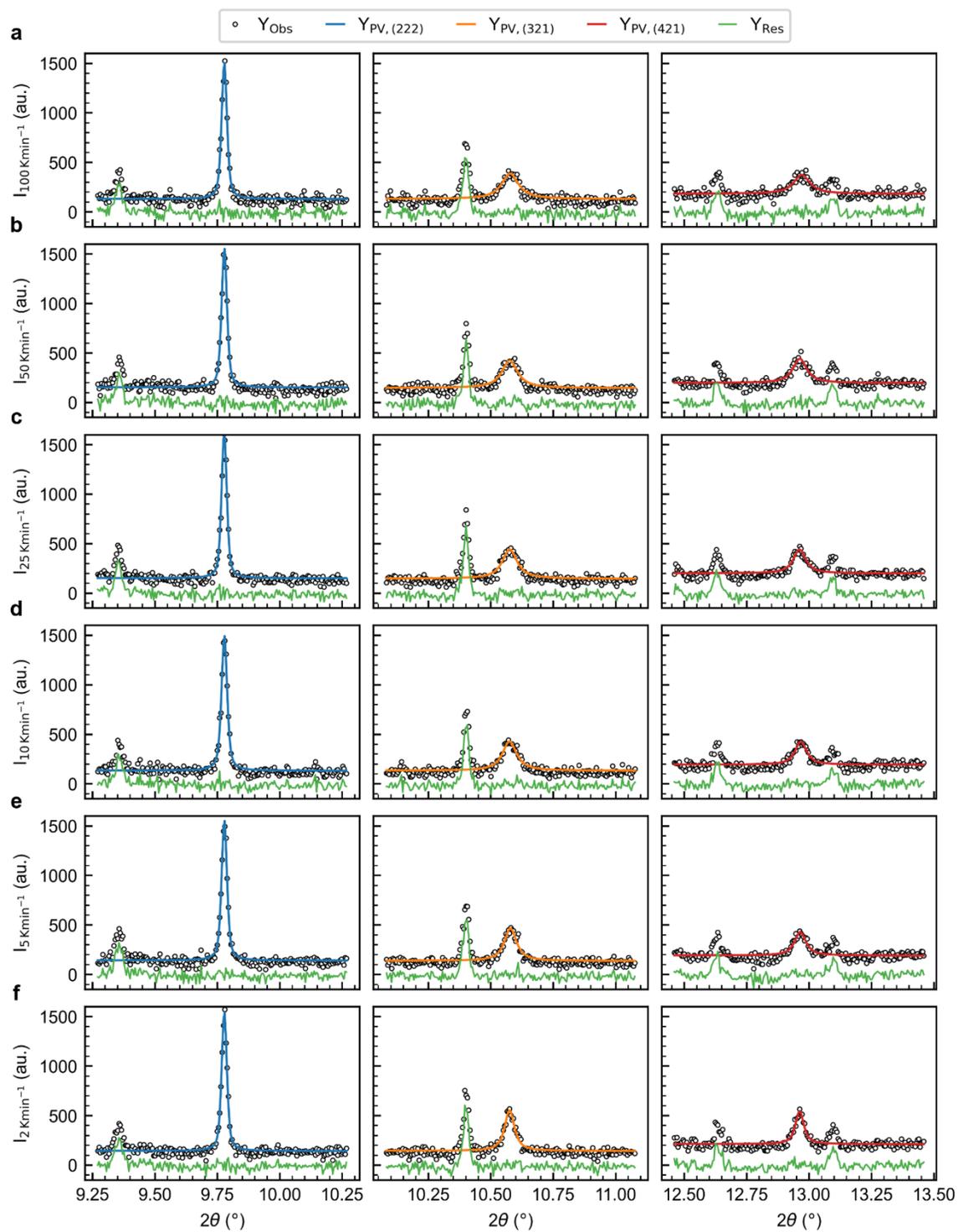

**Fig. S6. Single peak fitting at various cooling rates.** Single peak fitting of the (222) (blue), (321) (orange), and (421) (red) collected at 300 K using cooling rates of a) 100 Kmin$^{-1}$, b) 50 Kmin$^{-1}$, c) 25 Kmin$^{-1}$, d) 10 Kmin$^{-1}$, e) 5 Kmin$^{-1}$, and f) 2 Kmin$^{-1}$.

# 2 Transport properties

## 2.1 Temperature dependence of resistivity

Fig. S shows the effects of cooling rate on $\rho(T)$ curves, in which faster cooling rates result in poorer conductivity at low $T$'s. This cooling rate dependence is consistent with the VO domain picture described in the main text. The diamagnetism of Kutinaite also shows a dependence on cooling rate that is consistent with that of the resistivity (see Fig. S14).

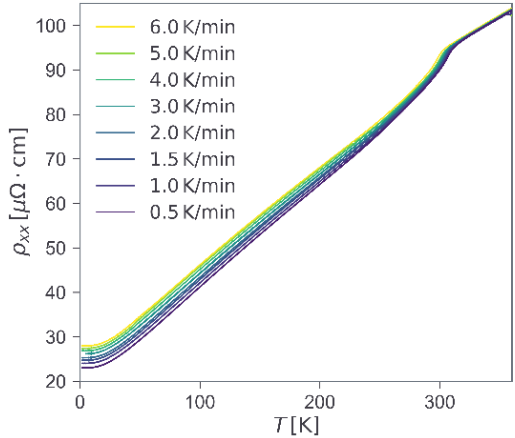

**Fig. S7. Dependence of cooling rate of the resistivity**

**Fig. S8. The hysteresis around $T_{VO}$.** Resistivity measured at different cooling and warming rates. Continuous lines and dashed lines show the cooling and warming up curves, respectively. Fig. S8 shows the hysteresis measured around the vacancy order transition at $T_{VO}$. The hysteresis expands with increasing the cooling/warming rate, and a similar phenomenon is also observed in the magnetization (Fig. S14).

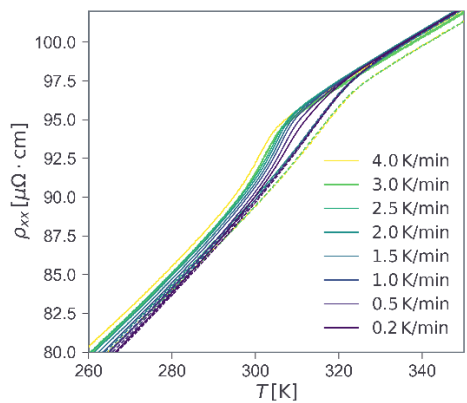

**Fig. S8. The hysteresis around $T_{VO}$.** Resistivity measured at different cooling and warming rates. Continuous lines and dashed lines show the cooling and warming up curves, respectively.

## 2.2 Magnetoresistance and Hall effect

The magnetic field ($H$) dependencies of magnetoresistance ($\Delta\rho$) and Hall resistivity ($\rho_{yx}$), shown in Fig. S9(a) and (b), suggest that multiple types of high mobility ($\mu$) charge carriers contribute to the conductivity of Kutinaite, even with the impeding effects of inter-grain scatterings inherent in sintered samples. At $T = 2$ K, $\Delta\rho$ is more than 11 % at 9 T. Increasing $T$ diminishes $\Delta\rho$, however it remains substantial at $T$ as high as 400 K.

Being consistent with $\Delta\rho$, at $T \leq 40$ K, the Hall resistivity ($\rho_{yx}(H)$) curves are also nonlinear with a negative slope bend at the low $H$-region and that turns to positive at $H > 6$ T. At $T > 100$ K, $\rho_{yx}(H)$ becomes linear in the whole $H$-window of $\pm 9$ T. Both $\Delta\rho$ and $\rho_{yx}(H)$ suggest a picture of parallel conductions of both electron-like and hole-like carrier types, the former is minor in number but has higher mobility. The magnetoresistive effect that continues up to 400 K suggests that the carrier mobilities are rather robust against elevated scattering rates at high temperatures.

### 2.1.1 Two-carrier-type analyses

The two carrier-type model and its low $H$ approximations[3], although being oversimplified, can be give a better insight of the magnetotransport of Kutinaite. The total conductivity is understood in terms of parallel contributions from two isotropic carrier species to the total conductivity of the material, $\sigma = \sigma_1 + \sigma_2 = e(n_1\mu_1 + n_2\mu_2)$, with $e$ is the (positive) value of the elemental charge, $n_i$ and $\mu_i$ are carrier number and mobility of the $i^{th}$ carrier species. If we use the convention that $n_i$ and $\mu_i$ are both negative (positive) for an electron-like (hole-like) carrier type, the formulae for the $H$- dependencies of Hall resistivity and magnetoresistance are:

$$\rho_{yx}(H) = \frac{H^3\mu_1^2\mu_2^2(n_1 + n_2) + H(\mu_1^2 n_1 + \mu_2^2 n_2)}{e(H^2\mu_1^2\mu_2^2(n_1 + n_2)^2 + (\mu_1 n_1 + \mu_2 n_2)^2)} \; ; \tag{S1}$$

$$\Delta\rho \equiv \frac{\rho_x(H)}{\rho_{xx}(0)} - 1 = \frac{H^2\mu_1\mu_2 n_1 n_2(\mu_1 - \mu_2)^2}{H^2\mu_1^2\mu_2^2(n_1 + n_2)^2 + (\mu_1 n_1 + \mu_2 n_2)^2} \; . \tag{S2}$$

Eq. (S1) for $\rho_{yx}(H)$, when combining with the conductivity $\sigma = 1/\rho$ at $H = 0$, is useful for estimating numbers and mobilities of both carrier types curves at $T \leq 40$ K, where the curvature of $\rho_{yx}(H)$ allows reliable fits. Fig. S9(a) shows that Eq. (S1) fits well to low $T$ $\rho_{yx}(H)$. **Error! Reference source not found.**Fig.5(c) and (d) in the main text summarize the results of the two-carrier-type Hall analysis, which indicates the coexistence of a fast electron-like carrier type which is minor in number ($\mu_1$ and $n_1$) and a major but slower hole-like carrier type ($\mu_2$ and $n_2$). At $T = 2$ K, $|\mu_1|$ is about $866 \, \text{cm}^2(\text{Vs})^{-1}$ and $|n_1| \approx 1.2 \times 10^{19} \, \text{cm}^{-3}$ in comparison with $\mu_2 \approx 100 \, \text{cm}^2(\text{Vs})^{-1}$ and $n_2 \approx 2.37 \times 10^{21} \, \text{cm}^{-3}$. However, plugging the $n_i$ and $\mu_i$ parameters obtained from the Hall analysis to (S2) results in a largely underestimated $\Delta\rho$, as evidenced in Fig. S9(b). Similar disagreements between Hall analysis and magnetoresistance are often the signatures of complex and/or nontrivial structures of Fermi surface[4-7].

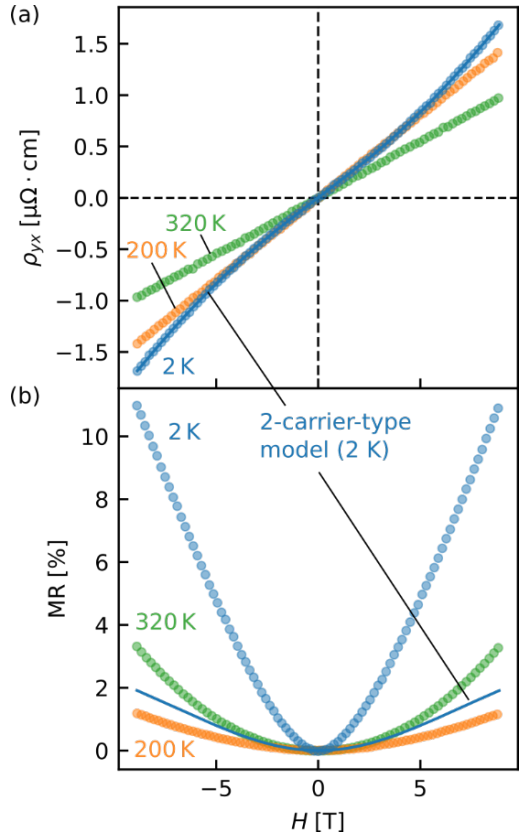

**Fig. S9. Hall effect and magnetoresistance.** (a) Two-carrier-type analysis of $\rho_{yx}(H)$. The fits are almost identical to the data. (b) A comparison between experimental $\Delta\rho(H)$ and calculations by substituting parameters obtained from the Hall analyses to Eq. (S1).

With increasing $T$, $\rho_{yx}(H)$ gradually flatten to linear lines and the analyses using two-carrier-type model become unreliable. On the other hand, the substantial $\Delta\rho(H)$ still demonstrates a manifestation of multiple carrier types up to $T = 400$ K. The low-$H$ expansions of (S1) and (S2) then become helpful in giving the trends of the $T$-dependcies of carrier number and mobility. Because no clear transitional features can be seen in the range $T \leq 300$ K, it is likely that the numbers of both carrier types do not change. We then make use of the ratio of the carrier numbers estimated at low $T$'s, $k = n_1/n_2 \approx -0.01$. The low-$H$ Hall effects can be written as,

$$\rho_{yx}(H) \approx R_H H \equiv \frac{H}{en_H}, \tag{S3}$$

Where the number of major carrier type dominates the Hall number, i.e., $n_2 \approx n_H$. We can also estimate independently the "average" mobility by a $H^2$-fit of $\Delta\rho$;

$$\Delta\rho(H) \approx \mu_{av}^2 H^2 \equiv [\alpha(|\mu_1| + \mu_2)]^2 H^2. \tag{S4}$$

Here $0 < \alpha = \sqrt{|k\mu_1/\mu_2|} < 1$. We applied these analyses in the extended the $T$-range from 2 K to 400 K. The results summarized in Fig. 5(c-d) of the main text show a rather featureless $n_H$ curve across the phase transition, which is a sharp contrast with the rich $T$-dependence of $\mu_\text{av}$. Initially diminished by increasing scattering rate at elevated $T$'s, $\mu_\text{av}$ exhibits a clear upward jump at $T \approx 280$ K and tends to enhance at higher $T$'s.

### 2.1.2 Quasi-linear magnetoresistance

$\Delta\rho$ at low $T$ exhibits a superlinear dependence on magnetic fields. Linear $\Delta\rho$ has been considered as a signature of Dirac fermions following Abrikosov model[4,5]. According to the model[5], $\Delta\rho$ is linear when the splitting $\Delta_1 = v_\text{F}(2\hbar eH)^{1/2}$ that the magnetic field $H$ opens between the zeroth and first Landau levels (LL) is larger than the Fermi energy $E_\text{F}$ and outweighs the thermal fluctuation $k_\text{B}T$. This means that all carriers can occupy only the zeroth LL at $H$ larger than a critical value $H^*$, which is,

$$H^* = \frac{1}{2e\hbar v_\text{F}^2}(k_\text{B}T + E_\text{F})^2. \tag{S5}$$

$H^*$ is the critical magnetic field at which $\Delta\rho'$ behavior crossovers from being parabolic (at $H < H^*$) to (super) linear (at $H > H^*$), and thus it corresponds to the turning point of the derivative $\partial\Delta\rho/\partial H$ as shown in Fig. S10(a). Using Eq.(S5), we then estimated that $E_\text{F} \approx 16$ meV and $v_\text{F} \approx 3.12 \times 10^5$ ms$^{-1}$ as Femi energy and Fermi velocity of pocket 1, respectively (Fig. S10(b)).

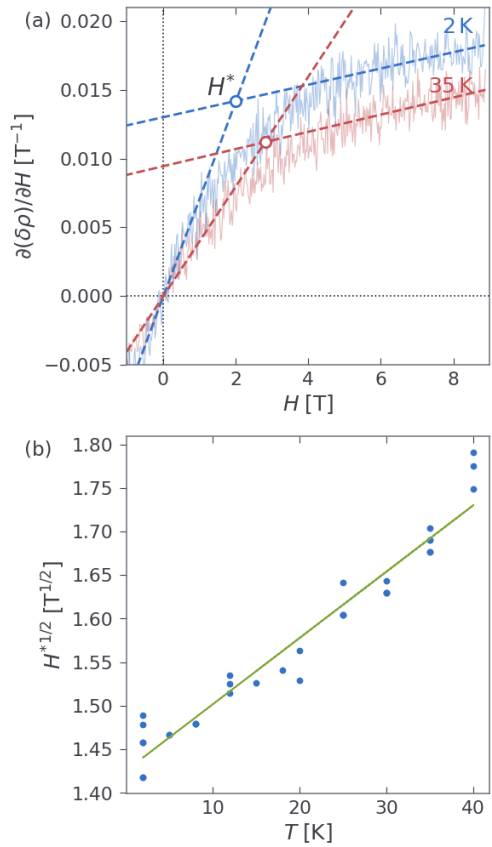

**Fig. S10. Modelling of magnetoresistance.** (a) $\partial(\Delta\rho)/\partial H$ at $T = 2$ K and 30 K and the estimations of $H^*$. (b) $T$-dependence of $H^*$ and with the fit by (S5).

## 3 Heat capacity

The specific heat capacity of Kutinaite shows a clear peak at $T_{VO}$ (Fig. S11(a)) and a large electronic contribution at low $T$'s (Fig. S11(b)). Fitting $C_p$ data at $T \leq 5$ K to $C_p/T = \gamma + AT^2$ yields $\gamma \approx 9$ mJ·mol$^{-1}$K$^{-2}$ as the electronic specific heat. Using the carrier number $n_2 \approx 2.38 \times 10^{21}$ cm$^{-3}$ estimated from the two-band analyses of Hall effect, the current $\gamma$ corresponds to effective mass ratio $m^*/m_e = \gamma/\gamma_e \approx 17$.

Plotting $C_p/T^3$ versus $T$ in the logarithmic scale exposes a large Boson-peak (Fig. S11(c)). Boson peak is the signature of excess vibrational density of states which do not obey long-range harmonic models. These vibrational states can come from anharmonic atomic motions, phonon-damping, or lattice soft modes. In our case, the boson peak is well-agreed with the anharmonic vibrations of the As2 atoms observed in SCXRD.

The Debye component in Fig. S11(c) was approximate as $C_{\text{Debye}} = 3RN \left(\frac{T}{\theta}\right)^3 \int_0^{x_D} dx \frac{x^4 \exp x}{(\exp x - 1)^2}$. Here $R$ and $N = 27.4$ are the ideal gas constant and the number of atoms per formula unit, respectively. $\theta$ is the Debye temperature estimated from the slope $A$ of the linear fit in Fig. S11(b), $\theta = (234NR/A)^{1/3}$, and $x_D = \theta/T$. The sole purpose of the model is to visualize the boson peak.

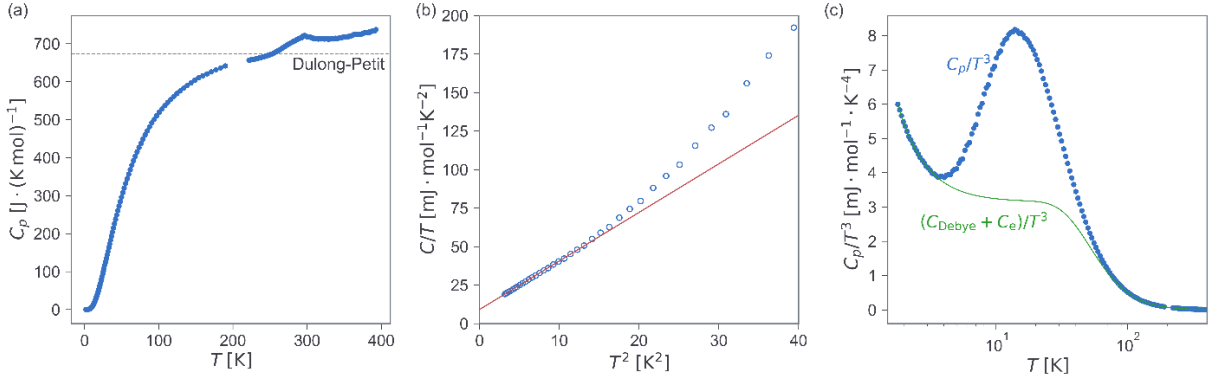

**Fig. S11. Specific heat.** (a) Specific heat of Kutinaite. (b) Extraction of electronic specific heat from the $C/T$ vs $T^2$ plot. (c) The Boson peak at low $T$'s.

## 4 Diamagnetism

We estimate the temperature dependence of the magnetic susceptibility $\chi$ using the magnetizations $M$ measured at 3 T and 5 T as follows.

$$\chi = \frac{M(5\,\mathrm{T}) - M(3\,\mathrm{T})}{2\,\mathrm{T}}. \tag{S6}$$

Fig. S12(a) shows the result in SI unit. We also calculate $\chi$ from the slopes of the linear fits of the isothermal magnetization curves measured at different temperatures (Fig. S12(b)). The value of $\chi$ resulted from method (squares in Fig. S12(a)) is consistent with the previous estimation using Eq. (S6). The small nonlinearity in the low field region of the $M(H)$ curves can arise from the small amount of impurities seen in PXRD.

Para- and diamagnetic components from various origins can be listed as follows.

$$\chi = \chi_\mathrm{P} + \chi_\mathrm{L} + \chi_\mathrm{cores} + \chi_\mathrm{R} + \chi_\mathrm{imp}. \tag{S7}$$

The first two terms, $\chi_\mathrm{P}$ and $\chi_\mathrm{L}$, have electronic origins. They account for paramagnetic Pauli and diamagnetic Landau behaviors of conducting electrons. On the other hand, $\chi_\mathrm{imp}$ is Curie-like paramagnetic contributions from impurities, which were removed using Eq. (S6). Other diamagnetism comes from intra-atomic (Lamor) or structurally bound ring currents as responses to applied magnetic field, i.e. $\chi_\mathrm{cores}$ and $\chi_\mathrm{R}$, respectively. Using standard atomic data[14], we estimated that $\chi_\mathrm{cores} \approx -1.19 \times 10^{-5}$ for Kutinaite. We cannot know $\chi_\mathrm{R}$ directly, however an upper limit of $-1.2 \times 10^{-5}$ can be taken from Clathrates[10]. Both $\chi_\mathrm{cores}$ and $\chi_\mathrm{R}$ are significantly smaller than the diamagnetic susceptibility observed experimentally. On the other hand, the high mobility metallic conduction suggests that Kutinaite's magnetic property is therefore dominantly defined by the electronic terms $\chi_P$ and $\chi_L$.

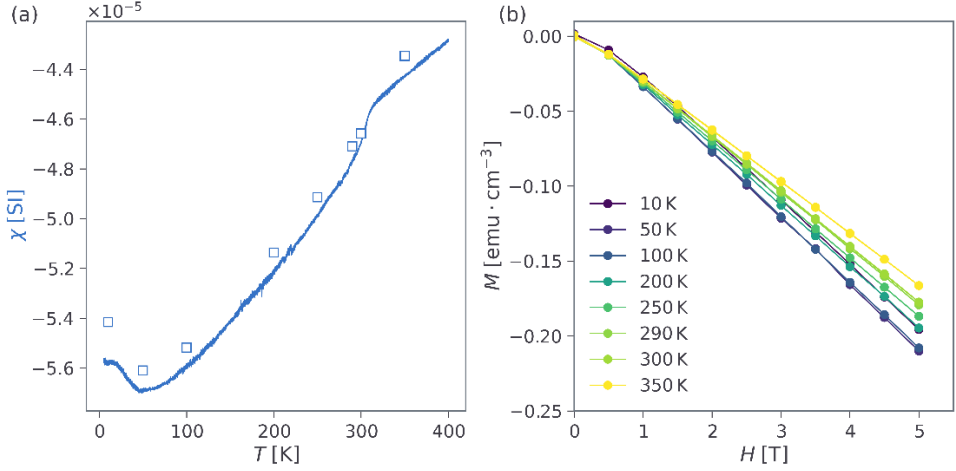

**Fig. S12. Magnetic properties.** (a) The diamagnetic susceptibility of Kutinaite (see text). (b) Isothermal magnetization curves measured at different temperatures.

Paramagnetic Pauli susceptibility is written as $\chi_P = \frac{1}{\hbar^2}\left(\frac{3}{\pi^4}\right)^{1/3}\mu_0\mu_B^2 m_e k n^{1/3}$ in the free electron model[15]. Here $\mu_0$ and $\mu_B$ are vacuum permeability and Bohr's magneton, respectively. $m_e$ is the mass of free electron, and $k = m^*/m_e$, with $m^*$ is the effective mass of electrons, and $n$ is the carrier number. On the other hand, Landau diamagnetism arises from orbital motions of conducting electrons, and its magnitude is inversely dependent on the effective mass $m^*$, i.e. $\chi_L = -\frac{1}{\hbar^2}\left(\frac{1}{3\pi^2}\right)^{2/3}\mu_0\mu_B \frac{m_e^2}{m^*} n^{1/3}$.

Using the carrier number $n_2 \approx 2.38 \times 10^{21}$ cm$^{-3}$ and $m^*/m_e = \gamma/\gamma_e \approx 17$ obtained from the analyses of Hall effect and specific heat yield $\chi_P \approx 7.6 \times 10^{-5}$. Considering all the components in Eq.(S7), we find that $\chi_L \approx -1.2 \times 10^{-4}$. Since microstrain effects in $C_p$ might lead to an overestimation of $\gamma$, we also make a more prudent calculation using $m^* = m_e$, which gives $\chi_P \approx 4.47 \times 10^{-6}$ and $\chi_L \approx -6.15 \times 10^{-5}$. Both estimations yield a Landau diamagnetism comparable with that of Bismuth [8,9] (see Table S4).

**Table S4**. Comparison of diamagnetic metals. The Landau diamagnetic susceptibility of Kutinaite is comparable with that of Bismuth.

| Material | $10^{-5}\chi$ [SI] | Ref. | |
|---|---|---|---|
| Kutinaite | −5.7 (total)<br>−12 (Landau) | This work | Metallic, LP |
| Bismuth | −17 | 8 | Metallic, LP |
| LaV$_2$Al$_{20}$ | −4.8 | 9 | Metallic, cage structure, LP |
| Ba$_6$Ge$_{25}$ | −1.2 | 10,11 | Weakly semiconducting, cage structure, RC |
| Graphite | −138 | 12,13 | Metallic, anisotropic LP and RC |
| Superconductors | −10$^5$ | - | - |

Fig. S13 displays the effect of cooling rate on χ. Slower cooling rates result in stronger diamagnetism. Because diamagnetic susceptibility originates from the cyclotron motions of electron-like carriers, reducing scattering will enhance its magnitude. As described in the main text, the formation of larger VO domain favorable at slower cooling rate leads to smaller fraction of domain wall states, and hence less scattering.

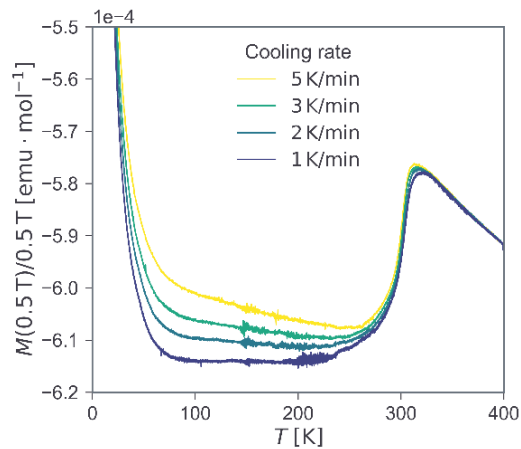

**Fig. S13. The effects of cooling rates on the diamagnetism of Kutinaite.** The magnetization was measured at $H = 0.5$ T.

Fig. S14 shows hysteresis around $T_{VO}$ in χ and the effects of cooling/warming rate. The hysteresis expands with increasing the cooling/warming rate.

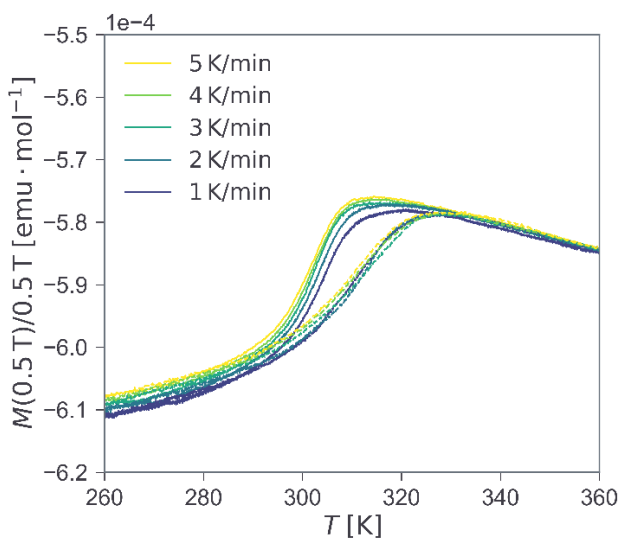

**Fig. S14. Hysteresis of magnetization around $T_{VO}$.** Magnetization measured at different cooling and warming rates. The applied magnetic field is 0.5 T. Continuous lines and dotted lines show the cooling and warming up curves, respectively.